\documentclass[aps,prb,twocolumn]{revtex4}
\usepackage{epsfig,amsmath,amssymb}
\newcommand{\be}{\begin{equation}}
\newcommand{\ee}{\end{equation}}
\newcommand{\ra}{\rangle}
\newcommand{\la}{\langle}
\newcommand{\bit}{\begin{itemize}}
\newcommand{\eit}{\end{itemize}}
\newcommand{\bea}{\begin{eqnarray}}
\newcommand{\eea}{\end{eqnarray}}

\newcommand{\Neel}{N\'{e}el }

\usepackage{epsfig}

\usepackage{color}

\begin{document}

\title
{Ground-state phases of the spin-$1/2$ $J_1$--$J_2$ Heisenberg antiferromagnet on the square lattice:
A high-order coupled cluster treatment}

\author
{R.~Darradi$^{1}$, O.~Derzhko$^{1,2}$, R.~Zinke$^{1}$, J.~Schulenburg$^{3}$, 
S.~E.~Kr\"uger$^{4}$, and J.~Richter$^{1}$}

\affiliation
{$^{1}$Institut f\"ur Theoretische Physik, Universit\"at Magdeburg,
P.O. Box 4120, 39016 Magdeburg, Germany
\\
$^{2}$Institute for Condensed Matter Physics, National Academy of Sciences of Ukraine,
1 Svientsitskii Street, L'viv-11, 79011, Ukraine
\\
$^{3}$Universit\"{a}tsrechenzentrum, Universit\"{a}t Magdeburg,
P.O. Box 4120, 39016 Magdeburg, Germany
\\
$^{4}$IESK, Kognitive Systeme, Universit\"at Magdeburg,
P.O. Box 4120, 39016 Magdeburg, Germany}

\date{\today}

\begin{abstract}
Using the coupled cluster method for high orders of approximation
and complementary exact diagonalization studies
we investigate the ground state properties
of the spin-$1/2$ $J_1$--$J_2$ frustrated Heisenberg antiferromagnet on the square lattice.
We have calculated the ground-state energy, the magnetic order parameter, the spin stiffness,
and several generalized susceptibilities to probe magnetically disordered
quantum valence-bond phases. We have found that the quantum critical points for both the N\'eel and collinear orders are
 $J_2^{c1}\approx (0.44 \pm 0.01)J_1$ and $J_2^{c2}\approx (0.59 \pm 0.01)J_1$  respectively, which are in good agreement with the results obtained by other approximations. 
In contrast to the recent study by [Sirker et al. Phys. Rev. B {\bf 73}, 184420 (2006)],
our data do not provide evidence
for the transition from the N\'{e}el to the valence-bond solid state
to be first order.
Moreover,
our results 
are in favor of 
the deconfinement scenario for that phase transition.
We also discuss the nature of the magnetically disordered quantum phase.
\end{abstract}

%PACS, the Physics and Astronomy Classification Scheme.
%Use showkeys class option if keyword display desired

\maketitle

\section{Introduction}
\label{intro}

Quantum phase transitions between semiclassical magnetically ordered phases
and magnetically disordered quantum phases
which are driven by frustration
attract much interest; see, e.g., Ref.~\onlinecite{Sachdev99}.
A canonical model for studying such transitions
is the spin-$1/2$ Heisenberg antiferromagnet
with nearest-neighbor $J_1$ and frustrating next-nearest-neighbor $J_2$ coupling
($J_1$--$J_2$ model) on the square lattice.
This model has attracted a great deal of interest during the last 20 years
(see, e.g.,
Refs.~\onlinecite{chandra88,dagotto89,figu90,read91,ivanov92,schulz,igarashi,richter93,richter94,dotsenko,
Einarsson95,zhito96,croo97,bishop98,singh99,capriotti00,capriotti01a,sushkov01,capriotti01,siu01,sushkov02,
capriotti03,singh03,roscilde04,zhang2003,Sir:2006,Mam:2006,Schm:2006,Scheidl:2006,Mune:2007}
and references therein).
Recent interest in this model comes also from the synthesis of layered magnetic materials
$\mathrm{Li}_{2}\mathrm{VOSiO}_4$,
$\mathrm{Li}_{2}\mathrm{VOGeO}_4$,
$\mathrm{VOMoO}_4$, 
and 
$\mathrm{BaCdVO(PO}_4)_2$ (Refs.~\cite{melzi00,rosner02,bombardi04,nath08})
that might be described by the $J_1$--$J_2$ model.
A new promising perspective is also opened 
by the recently discovered layered Fe-based superconducting materials \cite{kamihara} 
which may have a magnetic phase 
that can be described by a $J_1$--$J_2$ model with spin quantum number $s>1/2$.
\cite{yildirim,si,ma}

For the square-lattice spin-$1/2$ $J_1$--$J_2$ model it is well accepted
that there are two magnetically long-range ordered ground state (GS) phases at small and at large $J_2$
separated by an intermediate  quantum  paramagnetic phase without magnetic long-range order (LRO)
in the parameter region
$J_2^{c1} \le J_2 \le J_2^{c2}$,
where $J_2^{c1} \approx 0.4J_1$ and  $ J_2^{c2} \approx 0.6J_1$.
The magnetic phase at low $J_2 < J_2^{c1}$ exhibits semiclassical N\'eel LRO
with a magnetic wave vector ${\bf Q}_{0}=(\pi ,\pi )$.
The magnetic phase at large $J_2 > J_2^{c2} $ shows so-called collinear LRO.
It is twofold degenerate and the corresponding magnetic wave vectors are
${\bf Q}_{1} =(\pi , 0)$ or ${\bf Q}_{2} =(0 ,\pi )$. These two collinear states are characterized
by a parallel spin orientation of nearest neighbors in vertical (horizontal) direction
and an antiparallel spin orientation of  nearest neighbors in horizontal (vertical) direction.

The nature of the transition between the N\'eel and the quantum paramagnetic phases
as well as the properties of the quantum paramagnetic phase
and the precise values of the transition points, is still under debate. \cite{chandra88,dagotto89,figu90,read91,ivanov92,schulz,igarashi,richter93,richter94,dotsenko,
croo97,Einarsson95,zhito96,bishop98,singh99,capriotti00,capriotti01a,sushkov01,capriotti01,siu01,sushkov02,
capriotti03,singh03,roscilde04,zhang2003,Sir:2006,Mam:2006,Schm:2006,Scheidl:2006,Mune:2007}
In particular,
stimulated by the recent discussion of deconfined quantum criticality
in two-dimensional spin systems, \cite{Sent:2004a,Sent:2004b}
a renewed interest in the nature of the phase transition
between the semiclassical \Neel phase and the quantum paramagnetic phase
has emerged. \cite{Sir:2006,Scheidl:2006,gelle,kumar}
However, in spite of numerous intensive efforts 
focused on the transition between the N\'eel and the quantum paramagnetic phases 
in the $J_1$--$J_2$ square-lattice antiferromagnet
and some other candidate models, \cite{sandvik,kuklov,yoshioka,wenzel,alet,powell}
this field remains still highly controversial.

For completeness we mention that the classical square-lattice $J_1$--$J_2$ model
($s \to \infty$)
exhibits a direct first-order transition between N\'{e}el state and collinear state at 
$J_2/J_1=1/2$.

Recently, several extensions of $J_1$--$J_2$ model have been studied.
Interestingly, with increase in the space dimension from $D=2$ to $D=3$
the intermediate quantum paramagnetic phase disappears. \cite{schmidt02,oitmaa04,Schm:2006}
Also spatial \cite{Ne:2003,Si:2004,Star:2004,Mo:2006,bishop08}
and spin anisotropies \cite{roscilde04,viana07,bishop08a} as well as the spin
quantum number $s$ (Refs.~\onlinecite{chandra88,Scheidl:2006,bishop08b} and \onlinecite{bishop08c})
have a great influence on the GS phase diagram.

The goal of this paper is to study the GS phase diagram
for spin-half $J_1$--$J_2$ model on the square lattice
using a high-order coupled cluster method (CCM).
We complement the CCM treatment by exact diagonalization (ED) for 
finite lattices
for a qualitative check of our CCM data.
By calculating GS quantities such as
the energy,
the magnetic order parameter,
the spin stiffness
and generalized susceptibilities
we will investigate the quantum phase transitions present in the model
as well as  the properties of the quantum paramagnetic phase.
We will compare our results with the ones
obtained recently using series expansions. \cite{Sir:2006}

The CCM,
introduced many years ago by Coester and K\"ummel, \cite{coest}
is one of the most universal and most powerful methods of quantum many-body theory. For a review of the CCM see, e.g., Ref.~\onlinecite{bish_rev}.
Starting in 1990 it has been applied to quantum spin systems with much 
success. \cite{rog_her90,bishop91,bishop94,bishop98,
zeng98,krueger00,bishop00,krueger01,bishop04,rachid04,
farnell05,rachid05,Schm:2006,rachid06,zinke08}
A main advantage of this approach consists in its applicability
to strongly frustrated quantum spin systems in any dimension.
With the implementation of parallelization in the CCM code \cite{farnell05,cccm}
high-order calculations are now possible
(see Sec.~\ref{ccm}), improving significantly the accuracy in the investigation of quantum phase transitions driven by frustration.\cite{krueger00,krueger01,farnell05,rachid05,Schm:2006,zinke08,
bishop08,bishop08a}

The Hamiltonian of the considered $J_1$--$J_2$ model reads
\begin{eqnarray}
\label{ham}
H&=&J_1\sum_{\langle ij \rangle}{\bf s}_{i}{\bf s}_{j}
+J_2 \sum_{[ ij ]}{\bf s}_{i}{\bf s}_{j},
\end{eqnarray}
where $J_1$ is the nearest-neighbor exchange coupling and $J_2$ is the next-nearest-neighbor exchange coupling.
Both couplings are antiferromagnetic, $J_1>0$ and $J_2>0$.
In our CCM and ED calculations we set $J_1=1$.
We consider spin quantum number $s=1/2$, i.e., ${\bf{s}}_i^2=3/4$.

The remainder of the paper is organized as follows.
In Sec.~\ref{ccm} we briefly discuss the CCM approach
and illustrate how to calculate of GS quantities of spin model (\ref{ham}).
We present our results for the GS energy, the magnetic order parameter and
the spin stiffness in Sec.~\ref{ground_a}. In Sec.~\ref{ground_b}, we consider in more detail the phase transition between 
the N\'{e}el state and the quantum paramagnetic state
as well as various susceptibilities testing the nature of 
the nonmagnetic phase.
Finally,
in Sec.~\ref{sum} we summarize our findings.

\section{Coupled cluster method}
\label{ccm}

We start with a brief illustration of the main features of the CCM.
For a general  overview on the CCM the interested reader is referred,
e.g., to Refs.~\onlinecite{bishop91,zeng98,bishop00,krueger00,bishop04} and \onlinecite{farnell05,rachid05,rachid06}.
The starting point for a CCM calculation
is the choice of a normalized model (or reference) state $|\Phi\rangle$,
together with a set of mutually commuting multispin creation operators $C_I^+$
which are defined over a complete set of many-body configurations $I$.
The operators $C_I^-$ are the multispin destruction operators
and are defined to be the Hermitian adjoint of the $C_I^+$.
We choose $\{|\Phi\rangle;C_I^+\}$ in such a way
that we have $\langle\Phi|C_I^+=0=C_I^-|\Phi\rangle$, 
$\forall I\neq 0$.
Note that the CCM formalism corresponds to the thermodynamic limit $N\rightarrow\infty$.

For the spin system considered,
for $|\Phi\rangle$ we choose the two-sublattice \Neel state for small $J_2$
but the collinear state for large $J_2$.
To treat each site equivalently
we perform a rotation of the local axis of the spins such
that all spins in the reference state align along the negative $z$ axis.
In the rotated coordinate frame then we have
$|\Phi\rangle \hspace{-3pt} = \hspace{-3pt}
|\hspace{-3pt}\downarrow\rangle
|\hspace{-3pt}\downarrow\rangle
|\hspace{-3pt}\downarrow\rangle \ldots \,\,$
and the corresponding multispin creation operators then read
$C_I^+=s_i^+,\,\,s_i^+s_{j}^+,\,\, s_i^+s_{j}^+s_{k}^+,\cdots$,
where the indices $i,j,k,\dots$ denote arbitrary lattice sites.

The CCM parameterizations of the ket- and bra- ground states are given by
\begin{eqnarray}
\label{ket}
H|\Psi\rangle = E|\Psi\rangle ; 
\qquad  
\langle\tilde{\Psi}|H = E\langle\tilde{\Psi}| ;
\nonumber\\
|\Psi\rangle = e^S|\Phi\rangle, 
\qquad 
S = \sum_{I \neq 0}{\cal S}_IC_I^+ ; 
\nonumber\\
\langle\tilde{\Psi}| =  \langle\Phi|\tilde{S}e^{-S},
\qquad 
\tilde{S} = 1 + \sum_{I \neq 0}\tilde{\cal S}_IC_I^- .
\end{eqnarray}
The correlation operators $S$ and $\tilde {S}$
contain the correlation coefficients ${\cal S}_I$ and $\tilde {\cal S}_I$
that have to be determined.
Using the Schr\"odinger equation,
$H|\Psi\ra=E|\Psi\ra$,
we can now write the GS energy as $E=\la\Phi|e^{-S}He^S|\Phi\ra$.
The magnetic order parameter is given by
\be 
\label{mag}
M = -\frac{1}{N} \sum_{i=1}^N \la\tilde\Psi|s_i^z|\Psi\ra ,
\ee
where $s_i^z$ is expressed in the rotated coordinate system.
To find the ket-state and bra-state correlation coefficients
we require that the expectation value
$\bar H=\langle\tilde\Psi|H|\Psi\rangle$
is a minimum with respect to the bra-state and ket-state correlation coefficients,
such that the CCM ket- and bra-state equations are given by
\begin{eqnarray}
\label{ket_eq}
\langle\Phi|C_I^-e^{-S}He^S|\Phi\rangle = 0,
\qquad 
\forall I\neq 0,
\\
\label{bra_eq}
\langle\Phi|\tilde{\cal S}e^{-S}[H, C_I^+]e^S|\Phi\rangle = 0,
\qquad 
\forall I\neq 0.
\end{eqnarray}
Each ket- or bra-state [Eq.~\ref{ket_eq}) or ({\ref{bra_eq}}]
belongs to a particular index $I$ corresponding to a certain set
(configuration) of lattice sites
$i,j,k,\dots\;$ in the multispin creation operator
$C_I^+=s_i^+,\,\,s_i^+s_{j}^+,\,\, s_i^+s_{j}^+s_{k}^+,\cdots$; see above.

Though we start our CCM calculation with a reference state corresponding to semiclassical order,
one can compute the GS energy also in parameter regions where semiclassical magnetic LRO is destroyed,
and it is known \cite{bishop98,krueger00,rachid05,farnell05,Schm:2006,bishop08,bishop08a}
that the CCM yields precise results for the GS energy beyond the transition
from the semiclassical magnetic phase to the quantum paramagnetic phase.
The necessary condition for the convergence of the CCM equations
is a sufficient overlap between the reference state and the true ground state.

\begin{figure}%[ht]
\begin{center}
\epsfig{file=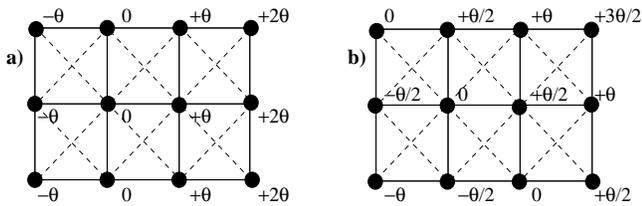,scale=0.55,angle=0.0}
\end{center}
\caption{Illustration of the twisted reference states
used for the calculation of the spin stiffness $\rho_s$.
The angles at the lattice sites indicate the twist of the spins
with respect to the \Neel or the collinear state.
(a): Twisted N\'{e}el state;
the twist is introduced along rows in $x$ direction.
(b): Twisted collinear state;
the twist is introduced along rows in $\vec{e}_x+\vec{e}_y$ direction.}
\label{fig01}
\end{figure}

It has been recently demonstrated \cite{rachid06}
that the CCM can also be used to calculate the spin stiffness ${\rho_s}$
with high accuracy.
The stiffness measures the increase in the amount of energy
when we twist the magnetic order parameter of a magnetically long-range ordered system
along a given direction by a small  angle $\theta$ per unit length,
i.e.
\be
\label{stiffn}
\frac{E(\theta)}{N} 
= 
\frac{E(\theta=0)}{N} + \frac{1}{2}\rho_s \theta^2 + {\cal O}(\theta^4),
\ee
where $E(\theta)$ is the GS energy as a function of the imposed twist,
and $N$ is the number of sites.
In the thermodynamic limit,
a positive value of $\rho_s$ means that there is magnetic LRO in the system,
while a value of zero reveals that there is no magnetic LRO.
To calculate the spin stiffness within the CCM using Eq.~(\ref{stiffn})
we must modify the corresponding reference states (N\'{e}el or collinear)
by introducing an appropriate twist $\theta$, see Fig.~\ref{fig01}.
Thus
the ket-state correlation coefficients ${\cal S}_I$
[after solving CCM equation (\ref{ket_eq})]
depend on $\theta$ and, hence, the GS energy $E$ is also dependent on $\theta$.

To study the properties of the quantum paramagnetic phase existing in the vicinity of $J_2=J_1/2$
as well as the phase transitions to that phase 
we will consider generalized susceptibilities $\chi_F$
that describe the response of the system to certain ''field'' operator $F$. \cite{capriotti00,capriotti01a,sushkov01,sushkov02,capriotti03,Sir:2006}
To calculate such a susceptibility $\chi_F$
we add to Hamiltonian (\ref{ham}) a field term $F=\delta\; \hat{O}$,
where $\hat{O}$ is an operator that breaks some symmetry of $H$
and the coefficient $\delta$ determines the strength of the field.
Using the CCM with either the \Neel or the collinear reference state
we calculate the energy per site
$E(\delta)/N=e(\delta)$
for $H+F$,
i.e., for the Hamiltonian of Eq.~(\ref{ham}) perturbed by the additional term $\delta \hat{O}$.
The susceptibility $\chi_F$ is then defined as
\begin{equation}
\label{suscept}
\chi_F = -\left.\frac{\partial^2{e(\delta)}}{\partial {\delta}^2} \right|_{\delta=0} \, .
\end{equation}

For the considered quantum spin model we have to use approximations
in order to truncate the expansion of $S$ and $\tilde {S}$.
We use the well elaborated LSUB$n$ scheme \cite{bishop91,zeng98,bishop00,bishop04,rachid05}
in which in the correlation operators $S$ and $\tilde {S}$
one takes into account all multispin correlations
over all distinct locales on the lattice defined by $n$ or fewer contiguous sites.
For instance,
within the LSUB4 approximation
one includes multispin creation operators of one, two, three or four spins
distributed on arbitrary clusters of four contiguous lattice sites.
The number of these fundamental configurations
can be reduced exploiting lattice symmetry and conservation laws.
In the CCM-LSUB10 approximation we have finally $29605$ ($45825$) 
fundamental configurations
for the \Neel (collinear) reference state.
Note, however, that for the calculation of the stiffness (the susceptibilities)
the twisted reference state (the modified Hamiltonian $H+F$) is less symmetric,
which leads to more fundamental configurations.
As a result we are then limited to LSUB8 approximation.

Since the LSUB$n$ approximation becomes exact for $n \to \infty$,
it is useful to extrapolate the ''raw'' LSUB$n$ data to $n \to \infty$.
Meanwhile there is a great deal of experience 
how to extrapolate the GS energy $e$ and the magnetic order parameter $M$. 
Most successful are the parameter fits of the form
$A(n)=A_0+A_1(1/n)^{\nu_1}+A_2(1/n)^{\nu_2}$ 
where the fixed leading exponents $\nu_1$ and $\nu_2$ may be different 
for the different quantities to be extrapolated.
For the GS energy per spin
$e(n) = a_0 + a_1(1/n)^2 + a_2(1/n)^4$
is a reasonable 
well-tested extrapolation ansatz. \cite{krueger00,bishop00,bishop04,rachid05,farnell05,
Schm:2006,bishop08,bishop08a}
An appropriate extrapolation rule for the magnetic order parameter 
for systems showing a GS order-disorder transition
is \cite{zinke08,bishop08,bishop08a,bishop08b}
$ M(n)=b_0+b_1(1/n)^{1/2}+b_2(1/n)^{3/2}$.
For the spin stiffness the extrapolation 
$\rho_s(n)=c_0+c_1 (1/n) + c_2(1/n)^2$ 
has been found to be reasonable. \cite{rachid06}
Finally,
for the susceptibility 
we have tested several fitting functions,
and we have found that the best extrapolation is obtained by the same 
fitting function
$\chi_F(n)=c_0+c_1 (1/n) + c_2 (1/n)^2$ 
as for the stiffness.
To check the reliability of this extrapolation scheme 
we have also performed an
extrapolation of the energy $e(\delta)$ to $n \to \infty $ by using the
extrapolation formula  
$e(n) = a_0 + a_1(1/n)^2 + a_2(1/n)^4\;$ 
(see above) 
and a subsequent calculation of $\chi_F$ according to Eq.~(\ref{suscept})
using the extrapolated energy.
We found that the deviations between both schemes are very small.

In summary,
the CCM approach automatically implies the thermodynamic limit $N\to\infty$ 
(that is an obvious advantage in comparison with ED). However, we need to extrapolate to the $n\to\infty$ limit in the truncation index $n$, 
which is an internal parameter of the approach. Since  
no general theory is known how the
physical quantities scale with $n$, we have to use extrapolation
formulas based on empirical experience. 
Another feature of many approximate techniques (but not of ED) 
is that they are based on reference states explicitly breaking some symmetry of the Hamiltonian.
Although CCM also starts from a reference state related to a particular magnetic LRO,
it has been demonstrated that the CCM provides precise results for the GS energy 
even in parameter regions where the magnetic LRO 
(i.e., the magnetic order parameter $M$ calculated within CCM)
vanishes. \cite{bishop98,krueger00,rachid05,farnell05,Schm:2006,bishop08,bishop08a}
This is again an advantage of the CCM approach. 

\section{Ground-state phase diagram}
\label{ground}

\subsection{Ground state energy, magnetic order parameter and spin stiffness}
\label{ground_a}

As already mentioned in Sec.~\ref{intro}, the considered $J_1$--$J_2$ model has two semiclassical magnetic GS phases 
(small and large $J_2$) 
separated by nonmagnetic quantum phase (intermediate $J_2$).
To detect the quantum critical points by the above described CCM 
we discuss the magnetic order parameter $M$  [see Eq.~(\ref{mag}], 
and the spin stiffness $\rho_s$ [see Eq.~(\ref{stiffn})]. 
Both, $M$ and $\rho_s$, are finite in the magnetically ordered phases 
but vanish in the intermediate quantum paramagnetic phase.

\begin{figure}%[!ht]
\begin{center}
\epsfig{file=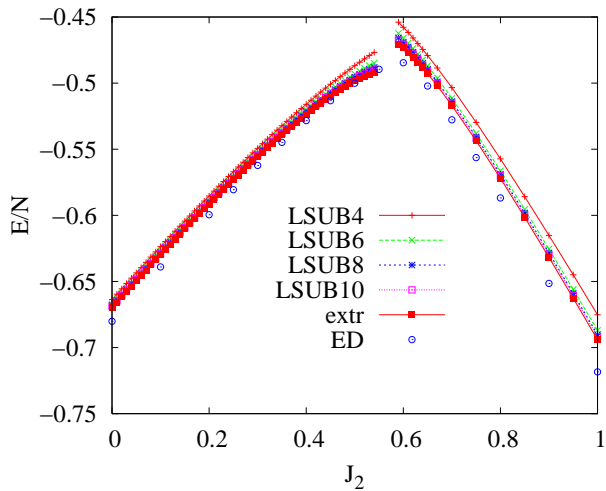,scale=0.75,angle=0.0}
\end{center}
\caption{(Color online)
The GS energy per spin as function of $J_2$ 
obtained by CCM-LSUB$n$ with $n=4, 6, 8, 10$ 
and its extrapolated values to $n \to \infty$
using the extrapolation scheme 
$e(n) = a_0 + a_1(1/n)^2 + a_2(1/n)^4$.
ED results for $N=32$ are shown by circles.
\label{fig02}}
\end{figure}

For completeness, we show first the CCM and the ED GS 
energies per spin, $e=E/N$, in Fig.~\ref{fig02}. 
%
%%%%%%%%%%%%%%%%%%%%%%%%%%%%%%%%%%%%%%%%%%%%%%%%%%%%%
%%%%%%%%%%%%%%%%%%%%%%%%%%%%%%%%%%%%%%%%%%%%%%%%%%%%%
%
The CCM curve consists of two parts 
corresponding to the \Neel and collinear reference states, respectively. 
The dependence $e(J_2)$ for ED and CCM is qualitatively the same; 
however,
due to finite-size effects, 
the ED curve is below the CCM curves.
Let us mention again that CCM GS energy 
corresponding to the N\'{e}el (collinear) reference state
is expected to be precise also in the intermediate quantum paramagnetic
phase if $J_2$ is not too far beyond the transition points.

\begin{figure}%[ht]
\begin{center}
\epsfig{file=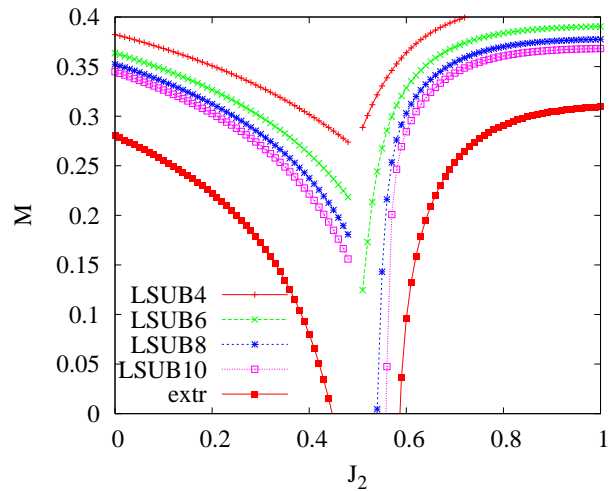,scale=0.75,angle=0.0}
\end{center}
\caption{(Color online)
Magnetic order parameter $M$ versus $J_2$ obtained by CCM-LSUB$n$ 
with $n=4, 6, 8, 10$ and its extrapolated values to $n \to \infty$
using the extrapolation scheme 
$M(n)=b_0+b_1(1/n)^{1/2}+b_2(1/n)^{3/2}$.}
\label{fig03}
\end{figure}

Next we consider the magnetic order parameter in dependence on $J_2$, 
see Fig.~\ref{fig03}. 
%
%%%%%%%%%%%%%%%%%%%%%%%%%%%%%%%%%%%%%%%%%%%%%%%%%%%%%
%%%%%%%%%%%%%%%%%%%%%%%%%%%%%%%%%%%%%%%%%%%%%%%%%%%%%
%
Note again that only for the magnetic order parameter $M$ and the GS
energy 
we are able to solve the CCM-LSUB$n$
equations up to $n=10$, while for the stiffness and the susceptibilities 
we are restricted to $n \le 8$. Hence the extrapolation to the limit 
$n \to \infty$ is most reliable for $M$ and the estimation of the phase
transition points by using the data for $M$ is most accurate.
The extrapolation to $n \to \infty $ shown in Fig.~\ref{fig03}
is based the 
extrapolation scheme $M(n)=b_0+b_1(1/n)^{1/2}+b_2(1/n)^{3/2}$ and uses 
CCM-LSUB$n$ data with $n=4, 6, 8, 10$. We find 
for the phase transition points between the semiclassical
phases and the quantum paramagnetic phase
$J_2^{c1} = 0.447J_1$ 
and
$J_2^{c2} = 0.586J_1$.
To check the robustness of this extrapolation 
we have also extrapolated $M$ 
using the data of  $n=2,4,6,8,10$ 
which leads to 
$J_2^{c1} = 0.443J_1$ 
and 
$J_2^{c2} = 0.586J_1$.
Those values $J_2^{c1}$ and $J_2^{c2}$ 
are in agreement with CCM predictions of Refs.~\onlinecite{bishop08} and \onlinecite{bishop08a}. 

Although the behavior of the extrapolated values of the magnetic order 
parameter around $J_2^{c1}$ and  $J_2^{c2}$
presented in Fig.~\ref{fig03}  shows a continuous behavior near $J_2^{c1}$ and  near $J_2^{c2}$, 
it is obvious that the decay of the collinear order parameter to zero at $J_2^{c2}$ is
much steeper than the decay of the \Neel order parameter at $J_2^{c1}$. 
That might 
give some hint of a first-order phase transition from the collinear to
the paramagnetic 
phase, in contrast to a continuous 
transition from the N\'{e}el to the paramagnetic phase.
\cite{schulz,singh99,sushkov01}

%
%{\sf{Ref.~\onlinecite{sushkov01}: The transition at $J_2^{c2}$ is probably of first order,
%but is very close to second order.}}

\begin{figure}%[!ht]
\begin{center}
\epsfig{file=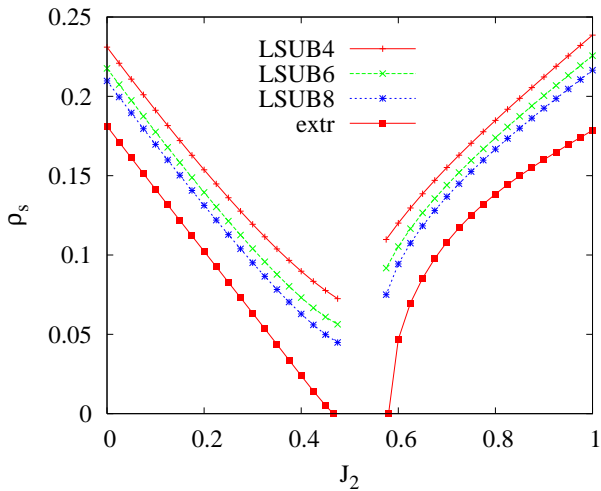,scale=0.75,angle=0.0}
\end{center}
\caption{(Color online)
The spin stiffness $\rho_s$ versus $J_2$ obtained by CCM-LSUB$n$ with
$n=4, 6, 8$ and its extrapolated values to $n \to \infty$
using the extrapolation scheme 
$\rho_s(n)=c_0+c_1(1/n)+c_2(1/n)^{2}$.}
\label{fig04}
\end{figure}

In addition to the magnetic order parameter,
another way to find the phase transition points 
is to consider the spin stiffness $\rho_s$
which is nonzero in a magnetically long-range ordered phase but vanishes in
the magnetically disordered quantum phase.  
The spin stiffness measures the distance of the ground state from criticality, \cite{chub94}
and constitutes together with the spin-wave velocity the fundamental parameters
that determines the low-energy dynamics of magnetic systems. \cite{Guida98,halperin69,chakra89}
In order to calculate the stiffness directly using Eq.~(\ref{stiffn})
we have to modify both the reference (N\'{e}el and collinear) states
by introducing an appropriate twist $\theta$; see Fig.~\ref{fig01}.
The CCM LSUB$n$ results for spin stiffness as well as the extrapolated values 
for both reference states as a function of $J_2$
are given in Fig.~\ref{fig04}.
%
%%%%%%%%%%%%%%%%%%%%%%%%%%%%%%%%%%%%%%%%%%%%%%%%%%%%
%%%%%%%%%%%%%%%%%%%%%%%%%%%%%%%%%%%%%%%%%%%%%%%%%%%%%%%%%%%%%%%%%%%%
%
The results show that approaching the magnetically disordered phase 
the stiffness
is decreased until it vanishes 
at $J_2=0.466 J_1$ coming from the \Neel phase 
and 
at $J_2=0.578 J_1$ coming from the collinear phase. 
These values obtained by extrapolation including up to LSUB8
data are in reasonable agreement with the critical points determined by
extrapolating $M$.
Note that our data for $\rho_s$ are also in good agreement 
with corresponding results of the other methods
(see Refs.~\onlinecite{ivanov92,Einarsson95,croo97} and \onlinecite{Trumper97}).
Note further that similarly as for $M$ we observe also for $\rho_s$ that the 
curvature near the critical
points is different at $J_2^{c1}$ and at $J_2^{c2}$ that might be again a
hint of the different nature of both transitions.

To summarize,
the CCM results for the GS energy, the magnetic order parameter, and the spin stiffness
support a general physical picture known from earlier numerical studies
(including ED, \cite{dagotto89,figu90,schulz,richter93,richter94}
variational quantum Monte Carlo, \cite{capriotti00,capriotti01} and series expansions \cite{sushkov02,Sir:2006}). For intermediate values of $J_2$,
$J_2^{c1}\le J_2\le J_2^{c2}$
with
 $J_2^{c1}\approx (0.44 \pm 0.01)J_1$ and $J_2^{c2}\approx (0.59 \pm 0.01)J_1$
%$J_2^{c1} \approx 0.44 \ldots 0.45J_1$ and $J_2^{c2}\approx
% 0.58 \ldots 0.59J_1$
there is no magnetic order.

\subsection{Order of the phase transition: Generalized susceptibilities}
\label{ground_b}

While the phase transition from the collinear to the paramagnetic phase 
is most likely of first order, \cite{schulz,singh99,sushkov01}
concerning the nature of phase transition from the N\'{e}el to the paramagnetic phase 
so far no conclusive answers are known. 
However,
the question about the order of the phase transition from the N\'{e}el to
the paramagnetic phase 
is of great interest
in particular in connection with the validity of the 
Landau-Ginzburg paradigm. \cite{Sent:2004a,Sent:2004b}
Very recently a number of arguments by Sirker et al. \cite{Sir:2006} based on series expansions and spin-wave theory were given
that this transition is of first order.
We reconsider this issue below using CCM and complementary ED results.

The first type of arguments in favor of the first-order phase transition from the N\'{e}el 
to the paramagnetic phase
presented in Ref.~\onlinecite{Sir:2006}
was based on the combination of field theory with series-expansion
data.
In what follows we use the same approach 
as that of Sirker et al.; \cite{Sir:2006}
however, instead of series-expansion
data we use CCM and ED data. 
Interestingly, we  
will arrive at a different conclusion concerning the nature of the phase
transition.

\begin{figure}%[!ht]
\begin{center}
\epsfig{file=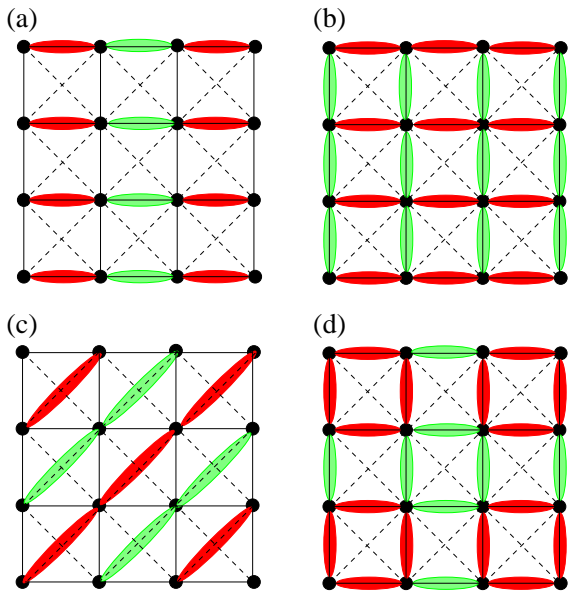,scale=0.51,angle=0.0}
\end{center}
\caption{(Color online)
Illustration of perturbations (fields) $F_j$ related to
generalized susceptibilities $\chi_j$: (a) perturbation $F_1$ (\ref{o1}), (b) perturbation $F_2$ (\ref{o2}), (c) perturbation $F_3$ (\ref{o3}) and (d) perturbation $F_4$ (\ref{o4}).
Dark (red) [light (green)] shadows correspond to enforced [weakened] exchange couplings.}
\label{fig05}
\end{figure}

The second type of arguments supporting the first-order phase transition 
from the N\'{e}el to the paramagnetic phase
were based on series-expansion data for several susceptibilities
that test a possible valence-bond solid (VBS) order in the paramagnetic phase.
In what follows we use the CCM and ED 
to compute four different susceptibilities $\chi_j$ defined in Eq.~(\ref{suscept}) 
for the  $J_1$--$J_2$ model.
The corresponding perturbations (fields) $F_j=\delta\; \hat{O}_j, \;
j=1,\ldots,4$,
are given by 
\be
\label{o1}
F_1 = \delta\; \sum_{i, j} (-1)^i{\bf s}_{i, j}{\bf s}_{i+1, j},
\ee
\be
\label{o2}
F_2 = \delta\;\sum_{i, j}\left({\bf s}_{i, j}{\bf s}_{i+1, j}-{\bf s}_{i, j}{\bf s}_{i, j+1}\right),
\ee
\be
\label{o3}
F_3 = \delta\;\sum_{i,j}(-1)^{i+j}
\left(s^x_{i,j}s^x_{i+1,j+1}+s^y_{i,j}s^y_{i+1,j+1}\right),
\ee
\be
\label{o4}
F_4 = \delta\;\sum_{i,j}
\left[(-1)^i{\bf s}_{i,j}{\bf s}_{i+1,j}+(-1)^j{\bf s}_{i,j}{\bf s}_{i,j+1}\right],
\ee
where $i,j$ are components (integer numbers) of the lattice vectors of the square lattice 
[see Fig.~\ref{fig05}), where we visualize perturbation terms (\ref{o1}) -- (\ref{o4})].
%
%%%%%%%%%%%%%%%%%%%%%%%%%%%%%%%%%%%%%%%%%%%%%%%%%%%%%%%%%%%
%%%%%%%%%%%%%%%%%%%%%%%%%%%%%%%%%%%%%%%%%%%%%%
%
%
The above definitions, Eqs.~(\ref{o1}) -- (\ref{o4}), are in accordance with
previous
discussions \cite{capriotti00,sushkov01,sushkov02,capriotti03} and \cite{Sir:2006}
of possible valence-bond states or broken symmetries in the magnetically disordered quantum phase.
Previous results for $\chi_1$ can be found in 
Refs.~\onlinecite{capriotti00,sushkov01,sushkov02,capriotti03} and \onlinecite{Sir:2006}, 
for $\chi_2$ in 
Refs.~\onlinecite{capriotti00} and \onlinecite{Sir:2006}, 
and for $\chi_3$ in Refs.~\onlinecite{sushkov02} and \onlinecite{Sir:2006}.
Note that in Refs.~\onlinecite{sushkov02} and \onlinecite{Sir:2006} 
the results for the perpendicular $\chi_3$ 
[i.e., the field $F_3=\delta\hat{O}_3$ contains only $x$ and $y$ components,
see Eq.~(\ref{o3})]
were reported only.
For reasons of comparison with the available series-expansion data
we consider 
in the present study also the perpendicular $\chi_3$.
To our best knowledge 
so far no data for the susceptibility $\chi_4$ are published.\cite{sushkov}

Note that all susceptibilities defined by Eqs.~(\ref{o1}) -- (\ref{o4}) 
break the symmetry of the initial square lattice, for
details, see
Refs.~\onlinecite{capriotti00,sushkov01,sushkov02,capriotti03} and \onlinecite{Sir:2006}. 
The susceptibilities $\chi_1$ and $\chi_4$ are most interesting, 
since they belong to order-parameter operators ${\hat O}_1$ and ${\hat O}_4$ 
probing directly possible valence-bond ordering. 
As discussed in Ref.~\onlinecite{Sent:2004b} 
they can also be interpreted as a single complex order parameter 
with a different phase for the two patterns. 
Note that for the field $F_1$ we have
chosen the $x$-axis for the alignment of modified nearest-neighbor
bonds, see Fig.~\ref{fig05}a. Alternatively, the $y$-axis can be chosen.    
It is worth mentioning that the field $F_4$ [Eq.~\ref{o4}] is a sum of 
fields $F_1$ aligned along $x$ and $y$ axes,
i.e., $F_4=F_1^{(x)}+F_1^{(y)}$,
and hence $\chi_4=\chi_1^{(x)}+\chi_1^{(y)}$.
If, in addition, a symmetry with respect to a $\pi/2$-rotation in the square-lattice plane holds
(that is, however, not the case, e.g., for the CCM calculations for large $J_2$),
one has $\chi_1^{(x)}=\chi_1^{(y)}$ and $\chi_4=2\chi_1$.

Analyzing the behavior of the susceptibilities as $J_2$ approaches the critical value $J_2^{c1}$
we will again arrive at a different conclusion in comparison to that in Ref.~\onlinecite{Sir:2006}.

\begin{figure}%[!ht]
\begin{center}
\epsfig{file=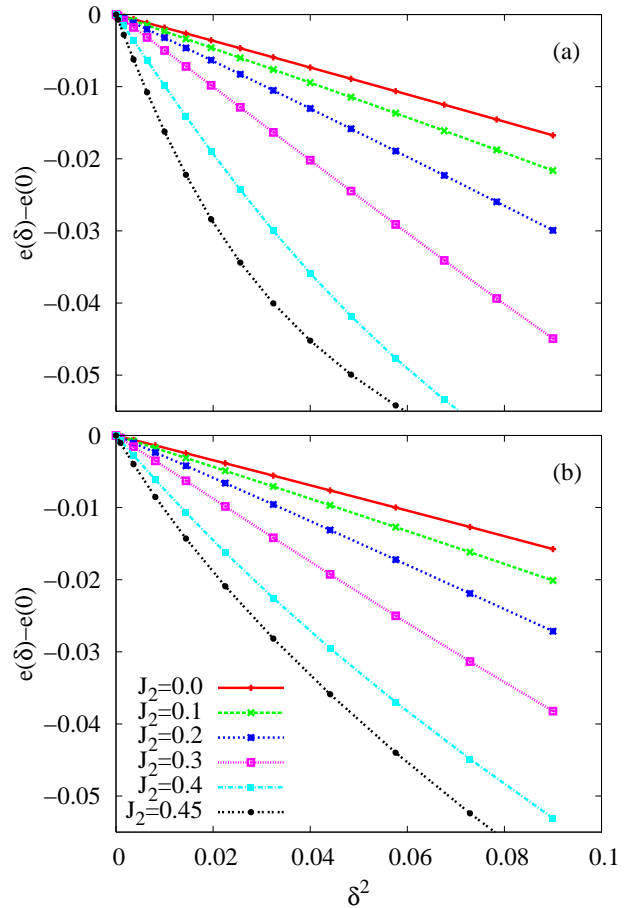,scale=0.75,angle=0.0}
\end{center}
\caption{(Color online)
The GS energy $e(\delta)-e(0)$ versus square of field strength $\delta$ for $H +
\delta {\hat O}_1$ [see Eq.~(\ref{o1}],
for $J_2=0.0,\;0.1,\;0.2,\;0.3,\;0.4$ and $0.45$ 
(from top to bottom). 
(a): CCM results extrapolated to $n\to\infty$ using the extrapolation scheme 
$e(n) = a_0 + a_1(1/n)^2 + a_2(1/n)^4$.
(b): ED results for $N=32$.
The displayed curves might be compared to the ones in Fig.~1 of Ref.~\onlinecite{Sir:2006}
where corresponding series-expansion  data for $e(\delta)$ 
are reported
(however, only up to $J_2=0.3$).}
\label{fig06}
\end{figure}

We begin with the examination of 
the order of the phase transition from the N\'{e}el to the VBS state. 
In contrast to the transition from the VBS to the collinear state 
where an energy level crossing indicates a first-order
transition, \cite{singh99,Sir:2006,bishop08} the energy behaves smoothly 
as $J_2$ varies around $J_2^{c1}$
%around $J_2 \sim J_2^{c1}$ 
and a more sensitive  method for distinguishing between first- and second-order transitions 
has to be applied. \cite{Sir:2006}
For that  we consider the GS energy $e(\delta)$ 
for Hamiltonian (\ref{ham}) perturbed by the field $F_1=\delta\hat{O}_1$ [Eq.~\ref{o1}].
We have performed CCM calculations for $e(\delta)$ 
choosing the \Neel state as the reference state
and 
extrapolating LSUB$n$ data with $n=4, 6, 8$ according to the scaling law 
$e(n) = a_0 + a_1(1/n)^2 + a_2(1/n)^4$
[see Fig.~\ref{fig06}a].
%
%%%%%%%%%%%%%%%%%%%%%%%%%%%%%%%%%%%%%%%%%%%%%%%%%%%%%%%%%%%
%%%%%%%%%%%%%%%%%%%%%%%%%%%%%%%%%%%%%%%%%%%%%%%%%%%%%%%%%%%
%
We have also performed complementary ED for a finite square lattice of $N=32$ sites
[see Fig.~\ref{fig06}b]
for a qualitative check of the CCM results.
The obtained dependence $e(\delta)$ may be fitted for a fixed $J_2$ to the following polynomial form
\begin{equation}
\label{Pol}
e(\delta)-e(0) = \frac{a}{2}\delta^2+\frac{b}{4}\delta^4+\frac{c}{6}\delta^6.
\end{equation}

To determine the order of the phase transition 
we use the method described in Ref.~\onlinecite{Sir:2006}.
For a  two-dimensional antiferromagnet,
the system could be described by the following $O(3)$ model:
\begin{equation}
\label{V_eq}
H_{v} = \frac{1}{2}\left[ (\partial_t \vec{v})^2 + c_v^2(\vec{\bigtriangledown}\vec{v})^2
+m_v^2\vec{v}^2\right] +\frac{u_v}{4}(\vec{v}^2)^2.
\end{equation}
Consider now the case that we are in the magnetically ordered phase
and add the field $F_1$ [Eq.~\ref{o1}] with $|\delta|\ll 1$.
The N\'eel order will then coexist with a small dimerization described by a scalar field
\begin{equation}
\label{S_eq}
H_\phi = \frac{1}{2}\left[ (\partial_t \phi)^2 + c_\phi^2(\vec{\bigtriangledown}\phi)^2
+m_\phi^2\phi^2\right] +\frac{u_\phi}{4}\phi^4+\frac{r_\phi}{6}\phi^6-\delta\phi.
\end{equation}
The fields $\vec{v}$ and $\phi$ are not independent,
and the interaction between them reads
\begin{equation}
\label{int_eq}
H_{\text{int}} = \frac{u_{v\phi}}{2}\vec{v}^2\phi^2.
\end{equation}
The effective field theory in the ordered phase for $\delta\neq 0$
is then given by $H=H_v+H_\phi+H_{\text{int}}$.
Combining Eqs.~(\ref{V_eq})-(\ref{int_eq}) we will have a nonzero GS expectation value
\begin{equation}
\label{VS_eq}
\langle\phi\rangle = \frac{\delta}{A}-\frac{u_\phi}{A^4}\delta^3+\frac{3u_\phi^2-Ar_\phi}{A^7}\delta^5+\mathcal{O}(\delta^7),
\end{equation}
with $A=m_\phi^2+u_{v\phi}\langle \vec{v}\rangle^2$.
Equation (\ref{VS_eq}) leads to a GS energy given by
\begin{equation}
\label{e_eq}
e(\delta)-e(\delta=0) = -\frac{1}{2A}\delta^2+\frac{u_\phi}{4A^4}\delta^4+\frac{Ar_\phi-3u_\phi^2}{6A^7}\delta^6+\mathcal{O}(\delta^8).
\end{equation}
The coefficient of the $\delta^4$ term in Eq.~(\ref{e_eq}) may be positive or negative
depending on the sign of the parameter $u_\phi$.
In the case of $u_\phi>0$ we have a second-order transition with respect to $\phi$ at a critical point,
and a first-order transition if $u_\phi<0$.

\begin{figure}%[!ht]
\begin{center}
\epsfig{file=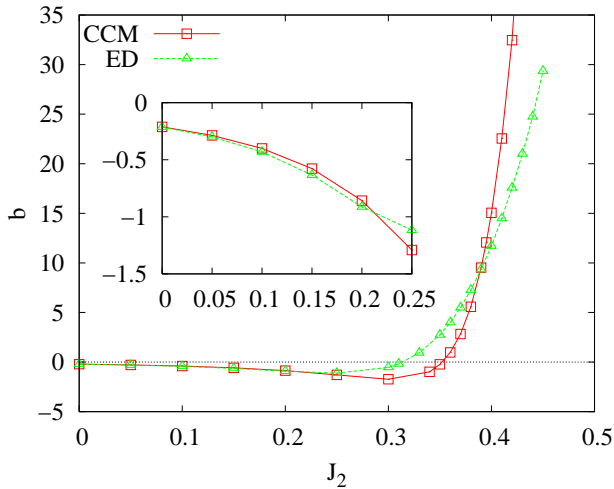,scale=0.75,angle=0.0}
\end{center}
\caption{(Color online)
The coefficient $b$ of the quartic term in Eq.~(\ref{Pol}) 
obtained from a fit of the CCM data in Fig.~\ref{fig06}a
and the ED data in Fig.~\ref{fig06}b in dependence on $J_2$.
This figure might be compared to Fig.~3 of Ref.~\onlinecite{Sir:2006}. 
Inset: the coefficient $b$ versus $J_2$ shown for small $J_2$ with an enlarged scale.}
\label{fig07}
\end{figure}

Using the polynomial in Eq.~(\ref{Pol})
we have fitted the data of $e(\delta)$, $\delta^2=0\ldots0.09$ 
for various $J_2$ including values near the critical point $J_2^{c1}$ 
[see Fig.~\ref{fig06}a].
We find that the coefficient of the $\delta^4$-term $b$ is negative for
small values of $J_2$ but becomes positive
if $J_2$ approaches $J_2^{c1}$; see Fig.~\ref{fig07}. 
%
%%%%%%%%%%%%%%%%%%%%%%%%%%%%%%%%%%%%%%%%%%%%%%%%%%%%%%%%%%%
%%%%%%%%%%%%%%%%%%%%%%%%%%%%%%%%%%%%%%%%%%%%%%%%%%%%%%%%%%%
%
This behavior is found for the CCM data as well as for the ED data. 
In particular,  $b$ calculated by the CCM (calculated by the ED) 
changes its sign  at $J_2 \approx 0.35$ (at $J_2 \approx 0.31$). 

Comparing Fig.~\ref{fig07} with the results reported in Fig.~3 of Ref.~\onlinecite{Sir:2006}
we note that CCM data for $J_2$ below 0.2 are in reasonable agreement 
with series expansions, linear spin-wave theory, or mean field spin-wave theory
[in particular, CCM yields 
$b(J_2=0.1)\approx -0.40$,
$b(J_2=0.2)\approx -0.86$,
$b(J_2=0.25)\approx -1.29$,
$b(J_2=0.3)\approx -1.73$
that is in between the series-expansion data and the spin-wave theory results].
A drastic difference between the series-expansion data and the CCM results emerges 
if $J_2$ approaches the critical value $J_2^{c1}$:
The series expansion gives $b<0$ whereas the CCM and the ED yield $b>0$ 
for $J_2 \to J_2^{c1}$.
We recall that any predictions from spin-wave theory for the considered $J_1$--$J_2$ model 
are likely to be unreliable if $J_2$ exceeds $0.35$. \cite{igarashi}
Combining Eqs. (\ref{Pol}) and (\ref{e_eq}) 
we get $b=u_{\phi}a^4$ 
and determining $a$ and $b$ using CCM data (Fig.~\ref{fig06}a) 
for $J_2=0.36\ldots0.42$ 
we find $u_{\phi}\approx 0.75>0$. 

In summary, the presented CCM and ED data, 
in contrast to series-expansion data of Ref.~\onlinecite{Sir:2006},
do not support a weak first-order phase transition from the N\'{e}el to the VBS state \cite{Sir:2006}
but give evidence that this transition is continuous.

\begin{figure}%[!ht]
\begin{center}
\epsfig{file=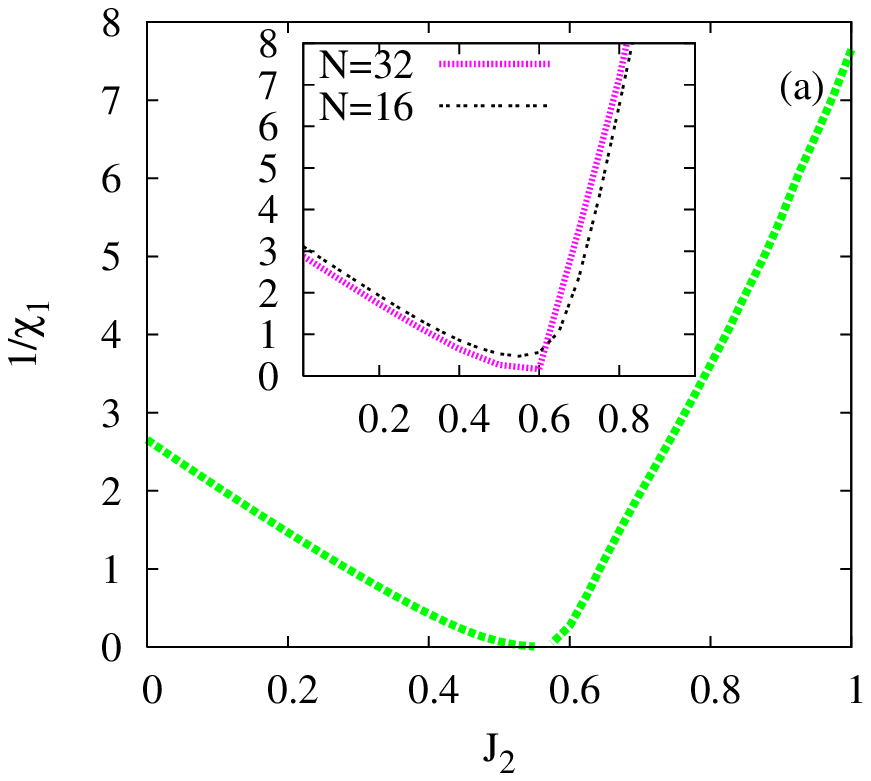,scale=0.71,angle=0.0}
\epsfig{file=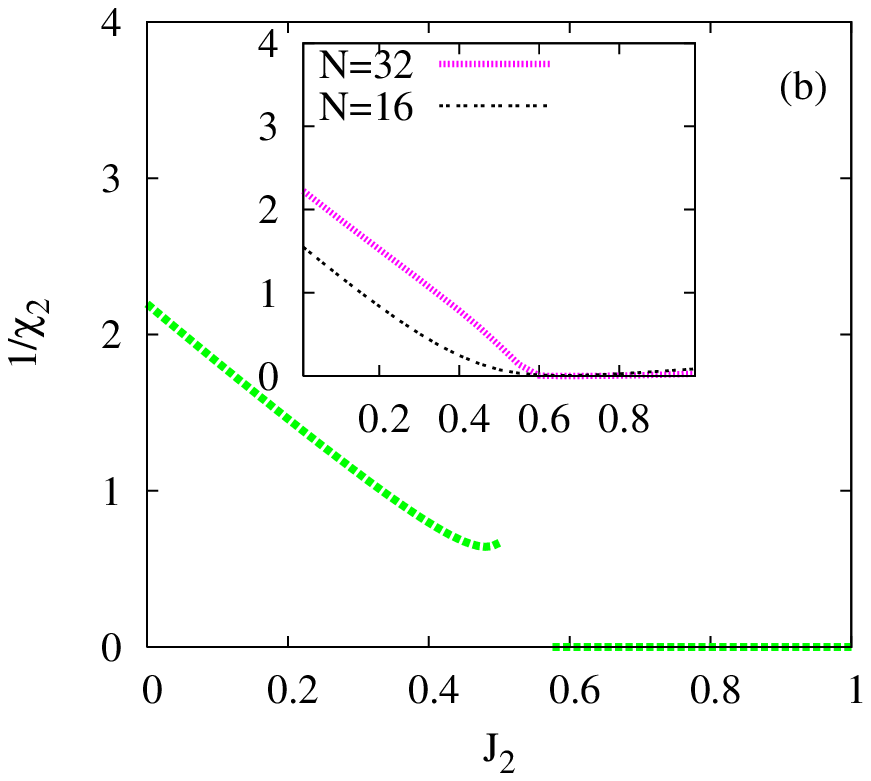,scale=0.71,angle=0.0}
\epsfig{file=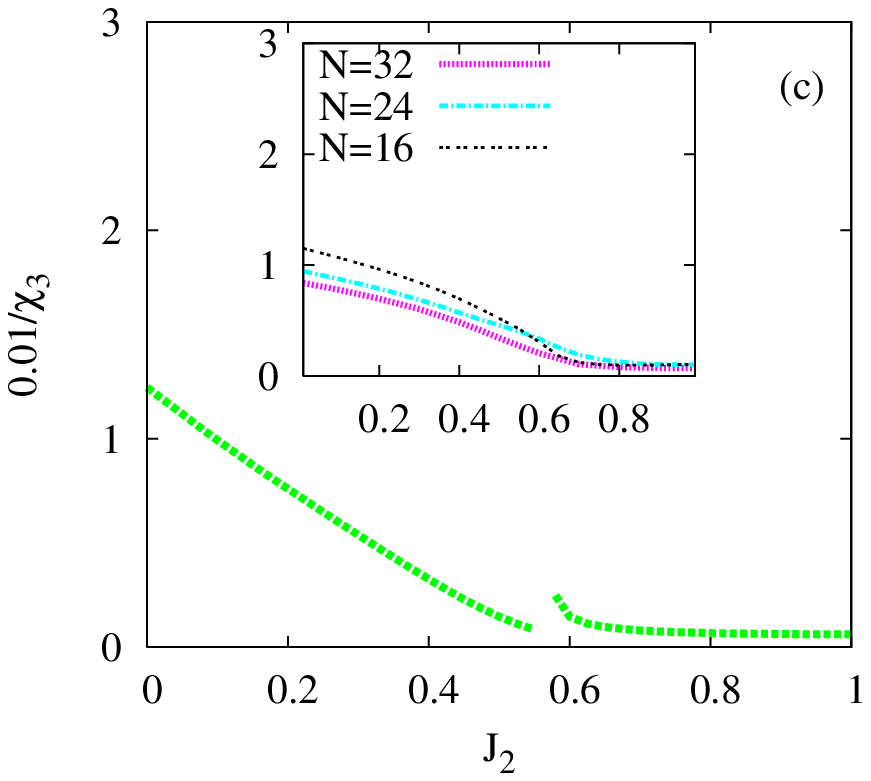,scale=0.71,angle=0.0}
\epsfig{file=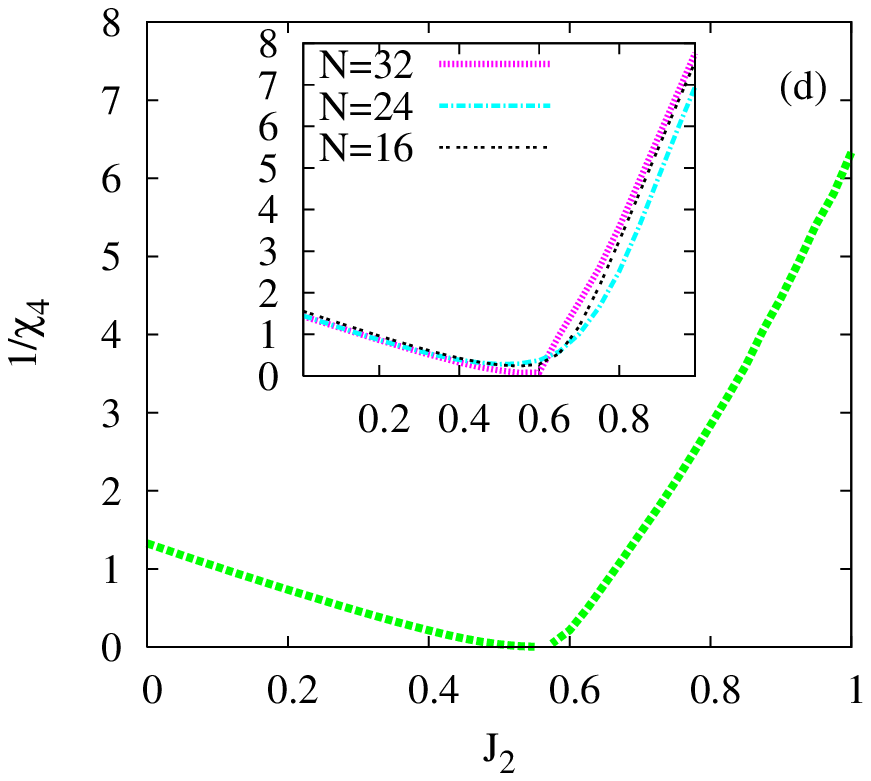,scale=0.71,angle=0.0}
\end{center}
\caption{(Color online)
The inverse susceptibilities 
(a) $1/\chi_1$, 
(b) $1/\chi_2$, 
(c) $1/\chi_3$
(please note the scaling factor $0.01$ at the $y$-axis), and  
(d) $1/\chi_4$ versus $J_2$
obtained within the CCM LSUB$n$ approximation with $n=4, 6, 8$
and extrapolated to $n\to\infty$ using $\chi(n)=c_0+c_1 (1/n) + c_2 (1/n)^2$.
Insets: the same as in the main panels but using ED for finite lattices of
$N=16,\;24$, and $32$.
Panel (a) might be compared to Fig.~2 of Ref.~\onlinecite{Sir:2006} 
and Fig.~3 of Ref.~\onlinecite{sushkov01},
panel (b) might be compared to Fig.~5 of Ref.~\onlinecite{Sir:2006}, and panel (c) might be compared to Fig.~6 of Ref.~\onlinecite{Sir:2006} 
and Fig.~3 of Ref.~\onlinecite{sushkov02}.}
\label{fig08}
\end{figure}

Next we examine the susceptibilities 
associated with probing fields (\ref{o1}) -- (\ref{o4}) directly.
The CCM results are shown in Fig.~\ref{fig08}.
%
%%%%%%%%%%%%%%%%%%%%%%%%%%%%%%%%%%%%%%%%%%%%%%%%%%%%%%%%%%%
%%%%%%%%%%%%%%%%%%%%%%%%%%%%%%%%%%%%%%%%%%%%%%%%%%%%%%%%%%%%
%
We also present in this figure the ED data for $N=16,\;24$, and $32$ lattice in the insets. (We do not show $N=24$ results for $\chi_1$ and 
$\chi_2$
since the system of rectangular shape perturbed by $F_1$ or $F_2$ does not possess symmetry 
with respect to a $\pi/2$ rotation in the square-lattice plane.) Note that a sophisticated finite-size analysis has to be performed 
in order to derive the correct behavior of susceptibilities in the thermodynamic limit. \cite{capriotti00}
Such an analysis goes beyond the scope of the present study since we use the ED data as a qualitative check of our CCM results only.
We notice here 
that although $\chi_1$ and $\chi_4$ are related to each other (see above),
they are calculated completely independently.
We have confirmed the expected relation between these susceptibilities
thus providing an additional double check for our numerics.

As it has been already mentioned above, the susceptibilities 
$\chi_1$, $\chi_2$, and $\chi_3$
were calculated in earlier studies 
using different methods.
Our CCM results for $\chi_1$ and $\chi_2$ are in a good quantitative agreement 
with series-expansion results reported for $J_2=0\ldots0.5$ 
in Refs.~\onlinecite{sushkov01} and \onlinecite{Sir:2006}. [For instance, one can compare our CCM data, 
$1/\chi_1(J_2=0.3) \approx 0.92$, 
$1/\chi_1(J_2=0.35)\approx 0.66$, 
and
$\chi_2(J_2=0.3)   \approx 0.90$, 
$\chi_2(J_2=0.35)  \approx 1.06$,
with the data shown in Figs.~2 and 5 of Ref.~\onlinecite{Sir:2006}.]
The CCM results for $\chi_1$ and $\chi_2$ also qualitatively agree 
with variational quantum Monte Carlo method and ED results reported (for some $J_2$
only) in Ref.~\onlinecite{capriotti00}.
The CCM results for  $\chi_3$, however, exhibit a different qualitative dependence on $J_2$ 
as $J_2$ approaches $J_2^{c1}$ in comparison with series-expansion data. \cite{sushkov02,Sir:2006}
Compare, e.g.,  Fig.~6 of Ref.~\onlinecite{Sir:2006}
and Fig.~\ref{fig08}c of the present paper.
According to series-expansion data 
$\chi_3$ decreases by about 20\% as $J_2$ increases from 0 to 0.4.
In contrast, according to CCM data shown in Fig.~\ref{fig08}c
$\chi_3$ increases by a factor of about 4 as $J_2$ increases from 0 to 0.4.

Let us now discuss some general features of the generalized susceptibilities 
shown in Fig.~\ref{fig08}.
Obviously, a divergence of a certain susceptibility (or $1/\chi\to 0$) 
at a particular value of $J_2$ 
indicates an instability of a GS phase regarding to a possible different GS order. 
It can be seen from Fig.~\ref{fig08}, that 
all susceptibilities increase with growing  $J_2$ in the \Neel phase. Near
the critical point  
$J_2^{c1}$ both $1/\chi_1$ and $1/\chi_4$ 
(CCM data imply $\chi_4=2\chi_1$)
are significantly smaller than 
$1/\chi_2$ and $1/\chi_3$, indicating 
that the valence-bond states belonging to the columnar dimerized and plaquette patterns 
are favorable in the magnetically disordered quantum phase.

A similar behavior of $\chi_1$ and $\chi_4$ 
(CCM data imply $\chi_4=\chi_1^{(x)}+\chi_1^{(y)}$)
is observed if $J_2$ approaches
$J_2^{c2}$ form the collinear phase, i.e., from $J_2 > J_2^{c2}$. 
On this side the behavior of $\chi_2$ and $\chi_3$ is not conclusive, 
since both are already large in the collinear phase.

\begin{figure}%[!ht]
\begin{center}
\epsfig{file=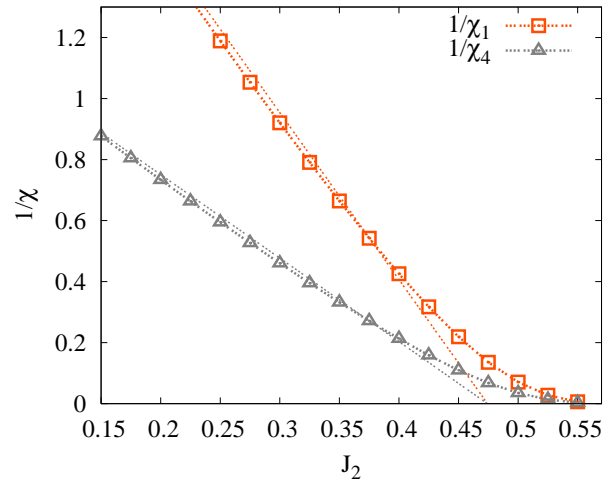,scale=0.75,angle=0.0}
\end{center}
\caption{(Color online)
Susceptibilities $1/\chi_1$ (red) and $1/\chi_4$ (gray) 
around the critical point $J_2^{c1}$.
Bold curves correspond to the CCM curves shown in Figs.~\ref{fig08}a and \ref{fig08}d.
Thin lines obtained from a linear fit of the CCM data for $0\le J_2\le
J_2^{c1} $.
Extrapolated (thin) lines become zero at $J_2\approx 0.47$.}
\label{fig09}
\end{figure}

The behavior of the susceptibilities $\chi_1$ and $\chi_4$ ($=2\chi_1$)
near the critical point $J_2^{c1}$ is shown in more detail in Fig.~\ref{fig09}.
%
%%%%%%%%%%%%%%%%%%%%%%%%%%%%%%%%%%%%%%%%%%%%%%%%%%%%%%%%%%%
%%%%%%%%%%%%%%%%%%%%%%%%%%%%%%%%%%%%%%%%%%%%%%%%%%%%%%%%%%%
%
Obviously, approaching $J_2^{c1}$ from the \Neel phase,
$\chi_1$ ($\chi_4$) becomes very large; it however, remains finite in the region around $J_2^{c1}$ up to $J_2=0.55$.
That might be attributed to limited 
accuracy of CCM results since: (i) we have data only up to LSUB8 for extrapolation 
and (ii) and LSUB$n$ data based on the N\'{e}el reference state 
may become less accurate for values of $J_2$ exceeding $J_2^{c1}$.
However, 
if the phase transition with respect to the corresponding VBS order parameter 
characterizing the quantum paramagnetic phase 
would be of second order we may expect an almost linear decrease in the inverse susceptibility 
if $J_2$ approaches $J_2^{c1}$, 
i.e., 
$1/\chi_1\propto (J_2^{c1}-J_2)^{\gamma_{\phi}}$ 
with 
$\gamma_{\phi}\approx 1$. \cite{Sir:2006} 
Hence a linear fit of the CCM data of $1/\chi_1$ ($1/\chi_4$)
versus $J_2$ 
using data points only within the \Neel ordered region 
$0\le J_2\le J_2^{c1}$ 
might give reasonable results.  
We find  that the linear fit for $1/\chi_1$ ($1/\chi_4$)
vanishes at the point $J_2\approx 0.47 J_1$; see Fig.~\ref{fig09}. 
This is in agreement with the scenario of deconfined criticality 
that predicts such divergence 
if the deconfined critical point is approached from the N\'{e}el phase. \cite{Sent:2004b,Sir:2006}

To conclude this part,
the CCM and ED data for all examined susceptibilities, 
$\chi_1$,  $\chi_2$,  $\chi_3$, and $\chi_4$,
exhibit an  enhancement 
while the system runs out of the N\'{e}el phase. 
This enhancement is most pronounced for $\chi_1$ ($\chi_4$).
Moreover, 
$\chi_1$ ($\chi_4$)  
diverges at a value of $J_2$ close to the quantum critical point
$J_2^{c1}\approx (0.44 \pm 0.01)J_1$ determined by the most accurate 
data for the \Neel order parameter $M$. This finding
is consistent with the predictions for a deconfined quantum critical point. \cite{Sent:2004b}
Furthermore we find that 
our CCM data for $\chi_1$ and  $\chi_2$ agree with the series-expansion data.\cite{Sir:2006} 
In contrast, for $\chi_3$ we observe a qualitatively different behavior.
Finally, the enhancement or divergence of the considered susceptibilities 
if $J_2$ approaches $J_2^{c1}$ 
indicates that the translational symmetry is broken in the quantum paramagnetic
phase, i.e. 
most likely a spatially homogeneous spin-liquid phase 
for $J_2^{c1} < J_2 < J_2^{c2}$ can be excluded.

\section{Summary}
\label{sum}

To summarize,
in this paper we have applied the CCM in high orders of approximation
to the spin-1/2 $J_1$--$J_2$ Heisenberg antiferromagnet on the square lattice 
and present a comprehensive analysis of the GS phase diagram of the model.
For this purpose 
we have calculated the GS energy, 
the magnetic order parameter, 
the spin stiffness 
and several generalized susceptibilities.
Our results enrich the list of available data 
and are complementary to other existing results 
obtained using different approximate methods such as series expansions 
or variational quantum Monte Carlo
for the spin-1/2 $J_1$--$J_2$ square-lattice Heisenberg antiferromagnet. In
addition to the CCM results we present also ED results that are found to be
in good agreement with the CCM data.

Our findings confirm the basic picture discussed earlier:
For intermediate values of $J_2^{c1}\le J_2\le J_2^{c2}$ the ground state is 
a paramagnetic quantum state. The 
CCM prediction for the boundaries of the paramagnetic region 
is 
 $J_2^{c1}\approx (0.44 \pm 0.01)J_1$ and $J_2^{c2}\approx (0.59 \pm
0.01)J_1$.
%$J_2^{c1}\approx 0.44 \ldots 0.45J_1$ 
%and 
%$J_2^{c2}\approx 0.58 \ldots 0.59J_1$.
To discuss the nature of the quantum phase transition from the semiclassical
\Neel phase to the quantum paramagnetic state at $J_2^{c1}$
we use the CCM (and ED) data as an input for the method developed  in 
Ref.~\onlinecite{Sir:2006}
to distinguish between a first- and a second-order transition. Our analysis  
leads to the conclusion 
that the phase transition from the N\'{e}el to the paramagnetic 
state at $J_2^{c1}$ is second order.
This outcome contradicts the conclusion of Ref.~\onlinecite{Sir:2006} 
based on series-expansion data,
but agrees with the deconfined critical point 
scenario proposed in Refs.~\onlinecite{Sent:2004a} and \onlinecite{Sent:2004b}.

Another way to check the predictions of the theory of deconfined quantum
criticality 
is to examine the susceptibilities related to  order parameters  of a possible
VBS ordering emerging,
if the critical point is approached from the magnetically ordered
\Neel phase.
The obtained data shown in Figs.~\ref{fig08} and \ref{fig09}
give another hint that $\chi_1$ ($\chi_4$) diverges at $J_2^{c1}$ 
which does not contradict the deconfined critical point scenario.
\cite{Sent:2004a,Sent:2004b}

Finally,
the divergence or enhancement of the generalized susceptibilities obtained by
CCM and ED
approaching $J_2^{c1}$ from the N\'{e}el phase
gives evidence in favor of ground states breaking translational symmetry. Therefore,
our data yield further arguments against a structureless 
(i.e., a spatially homogeneous)
spin-liquid state  without any LRO.

\section*{Acknowledgments}
We thank O. Sushkov for interesting discussions. The research was supported by the DFG 
(Projects No. Ri615/16-1 and No. Ri615/18-1).
O.D. acknowledges the kind hospitality of the University of Magdeburg 
in the spring of 2008.


\begin{thebibliography}{999}

\bibitem{Sachdev99}
S.~Sachdev, {\it Quantum Phase Transitions}
(Cambridge University Press, 1999);
S.~Sachdev, in {\it Quantum Magnetism}, Lecture Notes in Physics Vol.~{\bf 645}, edited by U.~Schollw\"ock, J.~Richter, D.~J.~J.~Farnell, and R.~F.~Bishop, 
(Springer, Berlin, 2004), p. 381.

\bibitem{chandra88}
P.~Chandra and B.~Doucot,
Phys. Rev. B {\bf 38}, 9335 (1988).

\bibitem{dagotto89}
E.~Dagotto and A.~Moreo,
Phys. Rev. Lett. {\bf 63}, 2148 (1989).

\bibitem{figu90}
F.~Figueirido, A.~Karlhede, S.~Kivelson, S.~Sondhi, M.~Rocek, and D.~S.~Rokhsar,
Phys. Rev. B {\bf 41}, 4619 (1990).

\bibitem{read91}
N.~Read and S.~Sachdev, 
Phys. Rev. Lett. {\bf 66}, 1773 (1991).

\bibitem{schulz}
H.~J.~Schulz and T.~A.~L.~Ziman,
Europhys. Lett. {\bf 18}, 355 (1992);
H.~J.~Schulz, T.~A.~L.~Ziman, and D.~Poilblanc,
J. Phys. I {\bf 6}, 675 (1996).

\bibitem{ivanov92}
N.~B.~Ivanov and P.~Ch.~Ivanov,
Phys. Rev. B {\bf 46}, 8206 (1992).

\bibitem{igarashi}
J.-i.~Igarashi,
J. Phys. Soc. Jpn. {\bf 62}, 4449 (1993).

\bibitem{richter93}
J.~Richter,
Phys. Rev. B {\bf 47}, 5794 (1993).

\bibitem{richter94}
J.~Richter, N.~B.~Ivanov, and K.~Retzlaff,
Europhys. Lett. {\bf 25}, 545 (1994).

\bibitem{dotsenko}
A.~V.~Dotsenko and O.~P.~Sushkov,
Phys. Rev. B {\bf 50}, 13821 (1994).

\bibitem{Einarsson95}
T.~Einarsson and H.~J.~Schulz,
Phys. Rev. B {\bf 51}, 6151 (1995).

\bibitem{zhito96}
M.~E.~Zhitomirsky and K.~Ueda,
Phys. Rev. B {\bf 54}, 9007 (1996).

\bibitem{croo97} 
M.~S.~L.~du~Croo~de~Jongh and P.~J.~H.~Denteneer, 
Phys. Rev. B {\bf 55}, 2713 (1997).

\bibitem{bishop98}
R.~F.~Bishop, D.~J.~J.~Farnell, and J.~B.~Parkinson,
Phys. Rev. B {\bf 58}, 6394 (1998).

\bibitem{singh99}
R.~R.~P.~Singh, Z.~Weihong, C.~J.~Hamer, and J.~Oitmaa,
Phys. Rev. B {\bf 60}, 7278 (1999).

\bibitem{capriotti00}
L.~Capriotti and S.~Sorella,
Phys. Rev. Lett. {\bf 84}, 3173 (2000).

\bibitem{capriotti01a}
L.~Capriotti,
Int. J. Mod. Phys. B {\bf 15}, 1799 (2001).

\bibitem{siu01}
L.~Siurakshina, D.~Ihle, and R.~Hayn,
Phys. Rev. B {\bf 64}, 104406 (2001).

\bibitem{sushkov01}
O.~P.~Sushkov, J.~Oitmaa, and Z.~Weihong,
Phys. Rev. B {\bf  63}, 104420 (2001).

\bibitem{capriotti01}
L.~Capriotti, F.~Becca, A.~Parola, and S.~Sorella,
Phys. Rev. Lett. {\bf 87}, 097201 (2001).

\bibitem{sushkov02}
O.~P.~Sushkov, J.~Oitmaa, and Z.~Weihong,
Phys. Rev. B {\bf  66}, 054401 (2002).

\bibitem{capriotti03}
L.~Capriotti, F.~Becca, A.~Parola, and S.~Sorella,
Phys. Rev. B {\bf 67}, 212402 (2003).

\bibitem{singh03}
R.~R.~P.~Singh, W.~Zheng, J.~Oitmaa, O.~P.~Sushkov, and C.~J.~Hamer,
Phys. Rev. Lett. {\bf 91}, 017201 (2003).

\bibitem{zhang2003} 
G.~M.~Zhang, H.~Hu, and L.~Yu, 
Phys. Rev. Lett. {\bf 91}, 067201 (2003).

\bibitem{roscilde04}
T.~Roscilde, A.~Feiguin, A.~L.~Chernyshev, S.~Liu, and S.~Haas,
Phys. Rev. Lett. {\bf 93}, 017203 (2004).

\bibitem{Sir:2006}
J.~Sirker, Z.~Weihong, O.~P.~Sushkov, and J.~Oitmaa,
Phys. Rev. B {\bf 73}, 184420 (2006).

\bibitem{Mam:2006}
M.~Mambrini, A.~L\"auchli, D.~Poilblanc, and F.~Mila,
Phys. Rev. B {\bf 74}, 144422 (2006).

\bibitem{Schm:2006}
D.~Schmalfu{\ss}, R.~Darradi, J.~Richter, J.~Schulenburg, and D.~Ihle,
Phys. Rev. Lett. {\bf 97}, 157201 (2006).

\bibitem{Scheidl:2006}
F.~Kr\"uger and S.~Scheidl,
Europhys. Lett. {\bf 74}, 896 (2006).

\bibitem{Mune:2007}
T.~Munehisa and Y.~Munehisa
J. Phys.: Condens. Matter {\bf 19}, 196202 (2007).

\bibitem{melzi00}
R.~Melzi, P.~Carretta, A.~Lascialfari, M.~Mambrini, M.~Troyer, P.~Millet, and F.~Mila,
Phys. Rev. Lett. {\bf 85}, 1318 (2000);
P.~Carretta, R.~Melzi, N.~Papinutto, and P.~Millet,
Phys. Rev. Lett. {\bf 88}, 047601 (2002);
P.~Carretta, N.~Papinutto, C.~B.~Azzoni, M.~C.~Mozzati, E.~Pavarini, S.~Gonthier, and P.~Millet,
Phys. Rev. B {\bf 66}, 094420 (2002).

\bibitem{rosner02}
H.~Rosner, R.~R.~P.~Singh, W. H.~Zheng, J.~Oitmaa, S.-L.~Drechsler, and W.~Pickett,
Phys. Rev. Lett. {\bf 88}, 186405 (2002).

\bibitem{bombardi04}
A.~Bombardi, J.~Rodriguez-Carvajal, S.~Di~Matteo, F.~de~Bergevin,
L.~Paolasini, P.~Carretta, P.~Millet, and R.~Caciuffo,
Phys. Rev. Lett. {\bf 93}, 027202 (2004).

\bibitem{nath08}
R.~Nath, A.~A.~Tsirlin, H.~Rosner, and C.~Geibel,
Phys. Rev. B {\bf 78}, 064422 (2008).
% arXiv:0803.3535v1 [cond-mat.str-el].
%BaCdVO(PO_4)_2

\bibitem{kamihara} 
Y.~Kamihara, T.~Watanabe, M.~Hirano, and H.~Hosono, 
J. Am. Chem. Soc. {\bf 130}, 3296 (2008).

\bibitem{yildirim}
T.~Yildirim, 
Phys. Rev. Lett. {\bf 101}, 057010 (2008).
% arXiv:0804.2252v1 [cond-mat.supr-con].

\bibitem{si}
Q.~Si and E.~Abrahams, Phys. Rev. Lett. {\bf 101}, 076401 (2008).

\bibitem{ma}
F.~Ma, Z.-Y.~Lu, and T.~Xiang, arXiv:0804.3370 (unpublished).

\bibitem{Sent:2004a}
T.~Senthil, A.~Vishwanath, L.~Balents, S.~Sachdev, and M.~P.~A.~Fisher,
Science {\bf 303}, 1490 (2004).

\bibitem{Sent:2004b}
T.~Senthil, L.~Balents, S.~Sachdev, A.~Vishwanath, and M.~P.~A.~Fisher,
Phys. Rev. B {\bf 70}, 144407 (2004).

\bibitem{gelle}
A.~Gell\'{e}, A.~M.~L\"{a}uchli, B.~Kumar, and F.~Mila,
Phys. Rev. B {\bf 77}, 014419 (2008).

\bibitem{kumar}
R.~Kumar and B.~Kumar,
Phys. Rev. B {\bf 77}, 144413 (2008).

\bibitem{sandvik}
A.~W.~Sandvik,
Phys. Rev. Lett. {\bf 98}, 227202 (2007).

\bibitem{kuklov}
A.~B.~Kuklov, M.~Matsumoto, N.~V.~Prokof'ev, B.~V.~Svistunov, and M.~Troyer,
Phys. Rev. Lett. {\bf 101}, 050405 (2008).
% arXiv:0805.4334v1 [cond-mat.stat-mech].

\bibitem{yoshioka}
D.~Yoshioka, G.~Arakawa, I.~Ichinose, and T.~Matsui,
Phys. Rev. B {\bf 70}, 174407 (2004).

\bibitem{wenzel}
S.~Wenzel, L.~Bogacz, and W.~Janke,
Phys. Rev. Lett. {\bf 101}, 127202 (2008).
%arXiv:0805.2500v1 [cond-mat.stat-mech].

\bibitem{alet}
F.~Alet, G.~Misguich, V.~Pasquier, R.~Moessner, and J.~L.~Jacobsen,
Phys. Rev. Lett. {\bf 97}, 030403 (2006).

\bibitem{powell}
S.~Powell and J.~T.~Chalker, Phys. Rev. Lett. {\bf 101}, 155702 (2008).

\bibitem{schmidt02}
R.~Schmidt, J.~Schulenburg, J.~Richter, and D.~D.~Betts,
Phys. Rev. B {\bf 66}, 224406 (2002).

\bibitem{oitmaa04}
J.~Oitmaa and W.~Zheng,
Phys. Rev. B {\bf 69}, 064416 (2004).

\bibitem{Ne:2003} 
A.~A.~Nersesyan and A.~M.~Tsvelik,
Phys. Rev. B {\bf 67}, 024422 (2003).

\bibitem{Si:2004}  
P.~Sindzingre,
Phys. Rev. B {\bf 69}, 094418 (2004).

\bibitem{Star:2004} 
O.~A.~Starykh and L.~Balents,
Phys. Rev. Lett. {\bf 93}, 127202 (2004).

\bibitem{Mo:2006} 
S.~Moukouri, J. Stat. Mech. (2006), P02002. 

\bibitem{bishop08}
R.~F.~Bishop, P.~H.~Y.~Li, R.~Darradi, and J.~Richter,
J. Phys.: Condens. Matter {\bf 20}, 255251 (2008).
% j1j1'j2 s=1/2

\bibitem{viana07} 
J.~R.~Viana and J.~R.~de~Sousa, 
Phys. Rev. B {\bf 75}, 052403 (2007).

\bibitem{bishop08a}
R.~F.~Bishop, P.~H.~Y.~Li, R.~Darradi, J.~Schulenburg, and J.~Richter,
Phys. Rev. B {\bf 78}, 054412 (2008).
%arXiv:0805.3107v1 [cond-mat.str-el].
% j1j2 s=1/2 xxz

\bibitem{bishop08b}
R.~F.~Bishop, P.~H.~Y.~Li, R.~Darradi, and J.~Richter,
Europhys. Lett. {\bf 83}, 47004 (2008).
% arXiv:0802.2566v1 [cond-mat.str-el].
% j1j1pj2 s=1

\bibitem{bishop08c}
R.~F.~Bishop, P.~H.~Y.~Li, R.~Darradi, J.~Richter, and C.~E.~Campbell,
J. Phys.: Condens. Matter {\bf 20}, 415213 (2008).
%arXiv:0805.4388v1 [cond-mat.str-el].
% j1j2 s=1 xxz

\bibitem{coest}
F.~Coester,
Nucl. Phys. {\bf 7}, 421 (1958); 
F.~Coester and H.~K\"ummel,
{\it ibid}. {\bf 17}, 477 (1960).

\bibitem{bish_rev}
R.~F.~Bishop,
in {\it Microscopic Quantum Many-Body Theories and Their Applications}, Lecture Notes in Physics Vol. {\bf 510}, edited by J.~Navarro and A.~Polls, (Springer, Berlin, 1998), p. 1.

\bibitem{rog_her90}
M.~Roger and J.~H.~Hetherington,
Phys. Rev. B {\bf 41}, 200 (1990).

\bibitem{bishop91}
R.~F.~Bishop, J.~B.~Parkinson, and Y.~Xian,
Phys. Rev. B {\bf 43}, 13782 (1991);
Phys. Rev. B {\bf 44}, 9425 (1991).

\bibitem{bishop94}
R.~F.~Bishop, R.~G.~Hale, and Y.~Xian,
Phys. Rev. Lett. {\bf 73}, 3157 (1994).

\bibitem{zeng98}
C.~Zeng, D.~J.~J.~Farnell, and R.~F.~Bishop,
J. Stat. Phys. {\bf 90}, 327 (1998).

\bibitem{krueger00} % square/honey
S.~E.~Kr\"uger,  J.~Richter, J.~Schulenburg, D.~J.~J.~Farnell, and R.~F.~Bishop,
Phys. Rev. B {\bf 61}, 14607 (2000).

\bibitem{bishop00} % square
R.~F.~Bishop, D.~J.~J.~Farnell, S.~E.~Kr\"uger,  J.~B.~Parkinson, and J.~Richter,
J. Phys.: Condens. Matter { \bf 12}, 6877 (2000).

\bibitem{krueger01} % square/honey
S.~E.~Kr\"uger and  J.~Richter,
Phys. Rev. B {\bf 64}, 024433 (2001).

\bibitem{bishop04}
D.~J.~J.~Farnell and R.~F.~Bishop,
in {\it Quantum Magnetism}, Lecture Notes in Physics Vol. {\bf 645}, edited by U.~Schollw\"ock, J.~Richter, D.~J.~J.~Farnell, and R.~F.~Bishop (Springer, Berlin, 2004), p. 307.

\bibitem{rachid04}
R.~Darradi, J.~Richter, and S.~E.~Kr\"uger,
J. Phys.: Condens. Matter {\bf 16}, 2681 (2004).

\bibitem{farnell05}
D.~J.~J.~Farnell,  J.~Schulenburg, J.~Richter, and K.~A.~Gernoth,
Phys. Rev. B {\bf 72}, 172408 (2005).

\bibitem{rachid05}
R.~Darradi, J.~Richter, and D.~J.~J.~Farnell,
Phys. Rev. B {\bf 72}, 104425 (2005).

\bibitem{rachid06}
S.~E.~Kr\"uger, R.~Darradi, J.~Richter, and D.~J.~J.~Farnell,
Phys. Rev. B {\bf 73}, 094404 (2006).

\bibitem{zinke08}
R.~Zinke, J.~Schulenburg, and J.~Richter,
Eur. Phys. J. B {\bf 64}, 147 (2008).

\bibitem{cccm}
For the numerical calculation we use the program package The CRYSTALLOGRAPHIC CCM (D.~J.~J.~Farnell and J.~Schulenburg).

\bibitem{chub94}
A.~V.~Chubukov, S.~Sachdev, and J.~Ye,
Phys. Rev. B {\bf 49}, 11919 (1994).

\bibitem{Guida98}
R.~Guida and J.~Zinn-Justin,
J. Phys. A {\bf 31}, 8103 (1998).

\bibitem{halperin69}
B.~I.~Halperin and P.~C.~Hohenberg,
Phys. Rev. {\bf 188}, 898 (1969).

\bibitem{chakra89}
S.~Chakravarty, B.~I.~Halperin, and D.~R.~Nelson,
Phys. Rev. B {\bf 39}, 2344 (1989).

\bibitem{Trumper97}
A.~E.~Trumper, L.~O.~Manuel, C.~J.~Gazza, and H.~A.~Ceccatto,
Phys. Rev. Lett. {\bf 78}, 2216 (1997);
L.~O.~Manuel, A.~E.~Trumper, and H.~A.~Ceccatto,
Phys. Rev. B {\bf 57}, 8348 (1998).
% stiffness mit schwinger Boson

%\bibitem{lnp04}
%{\it Quantum Magnetism},
%eds. U.~Schollw\"{o}ck, J.~Richter, D.~J.~J.~Farnell, and R.~F.~Bishop,
%Lecture Notes in Physics {\bf 645} (Springer, Berlin, 2004).

\bibitem{sushkov} Note, however, that in Ref.~\onlinecite{sushkov01}
the response to a plaquette-type modulation of bonds 
starting from a columnar
dimerized state was calculated using series expansion.  
\end{thebibliography}
\end{document}